\documentclass{natureprintstyle}

\usepackage{psfig}

\usepackage{lscape}
\usepackage{amsmath}
\usepackage{amssymb}
\usepackage{color}
\usepackage{url}
\usepackage{ulem}
\usepackage{multirow}
\usepackage{graphics,graphicx}
\usepackage[affil-it]{authblk}
\usepackage{blindtext}
\usepackage{subfig}
\usepackage{hyperref}

\usepackage{astjnlabbrev-nature}

\def\micron{$\mu$m}

\def\arcsec{''}

\def\spose#1{\hbox to 0pt{#1\hss}}
\def\simlt{\mathrel{\spose{\lower 3pt\hbox{$\mathchar"218$}}
     \raise 2.0pt\hbox{$\mathchar"13C$}}}
\def\simgt{\mathrel{\spose{\lower 3pt\hbox{$\mathchar"218$}}
     \raise 2.0pt\hbox{$\mathchar"13E$}}}

\makeatletter
\renewcommand{\section}{\@startsection%
{section}{1}{0mm}{-\baselineskip}%
{0.5\baselineskip}{\normalfont\Large\bfseries}}%
\makeatother

\title{Nuclear obscuration in active galactic nuclei}

\author{Cristina Ramos Almeida$^{1,2}$ \& Claudio Ricci$^{3,4,5}$\thanks{*The authors' order is purely alphabetical since they both have contributed equally to the Review. 
e-mail: cra@iac.es and cricci@astro.puc.cl}}

\begin{document}
\pagestyle{plain}
\pagenumbering{arabic}

\twocolumn[
  \begin{@twocolumnfalse}

\let\newpage\relax\maketitle

\begin{affiliations}
 \item Instituto de Astrof\' isica de Canarias, Calle V\' ia L\'actea, s/n, E-38205, La Laguna, Tenerife, Spain.
 \item Departamento de Astrof\' isica, Universidad de La Laguna, E-38206, Tenerife, Spain.
 \item Instituto de Astrof\' isica, Facultad de F\' isica, Pontificia Universidad Cat\'olica de Chile, Casilla 306, Santiago 22, Chile.
 \item Kavli Institute for Astronomy and Astrophysics, Peking University, Beijing 100871, China.
 \item Chinese Academy of Sciences South America Center for Astronomy and China-Chile Joint Center for Astronomy, Camino El Observatorio 1515, Las Condes, Santiago, Chile.
\end{affiliations}

%

\bigskip

    \begin{abstract}
        The material surrounding accreting supermassive black holes connects the active galactic nucleus (AGN) with its host galaxy and, besides being responsible for feeding the black hole, 
       provides important information on the feedback that nuclear activity produces on the galaxy. In this Review we summarize our current understanding of the close environment of accreting 
       supermassive black holes obtained from studies of local AGN carried out in the infrared and X-ray band. The structure of this circumnuclear material is complex, clumpy and dynamical, 
       and its covering factor depends on the accretion properties of the AGN. From the infrared point of view, this obscuring material is a transition zone between the broad- and 
       narrow-line region, and at least in some galaxies, it consists of two structures: an equatorial disk/torus and a polar component. In the X-ray regime, the obscuration is produced 
       by multiple absorbers on various spatial scales, mostly associated with the torus and the broad-line region. In the next decade the new generation of infrared and X-ray facilities will 
       greatly contribute to our understanding of the structure and physical properties of nuclear obscuration in AGN.
      \bigskip
    \end{abstract}
  \end{@twocolumnfalse}
  ]
{
  \renewcommand{\thefootnote}%
    {\fnsymbol{footnote}}
  \footnotetext[1]{\small The authors' order is purely alphabetical since they both have contributed equally to this Review. 
  
  e-mail: cra@iac.es and cricci@astro.puc.cl} 
}

Over the past decades several pieces of observational evidence have shown that supermassive black holes (SMBHs; $M_{\rm\,BH}\sim 10^{6-9.5}M_{\odot}$) 
are found at the center of almost all massive galaxies, and that those SMBHs play an important role in the evolution of their host galaxies\cite{Kormendy:2013uf} 
during a phase in which they are accreting material and are observed as active galactic nuclei (AGN). Indeed, different modes of AGN feedback are expected to be 
key processes shaping the environment of SMBHs. In particular, quasar-induced outflows might be capable of regulating black hole and galaxy growth\cite{DiMatteo05}. 
For instance, they are required by semi-analytical models of galaxy formation for quenching star formation in massive galaxies\cite{Croton06}. However, directly 
studying the influence of nuclear activity on galaxy evolution is difficult because of the completely different timescales involved\cite{Hickox14,Schawinski:2015cs}. 
{Therefore, to directly probe the AGN--host galaxy connection we need to look at the structure and kinematics of the parsec-scale dust and gas surrounding the 
accreting SMBHs.}


AGN radiate across the entire electromagnetic spectrum, from the radio {and up to gamma-rays}. A large fraction of their emission is produced in the accretion disk and emitted in the optical and ultraviolet (UV) bands.
 {A significant proportion of these optical/UV photons are reprocessed i) by dust located beyond the sublimation radius and re-emitted in the infrared (IR), and ii) by a corona of hot electrons close to the accretion disk that up-scatters them in the X-ray band\cite{Haardt:1994bq} and illuminates the surrounding material. Thus, studying the IR and X-ray emission and absorption of AGN is key to characterize nuclear regions of AGN.} 


\section*{AGN structure}



AGN are classified as type-1 and type-2 depending on the presence or not of broad components (full width at half maximum; FWHM$\geq$2000 km~s$^{-1}$) in the permitted lines of their optical spectra. Those broad lines are produced in a sub-pc scale dust-free region known as the broad-line region (BLR). On the other hand the narrow lines (FWHM$<$1000 km~s$^{-1}$) that are ubiquitous in the spectra of AGN --excluding beamed AGN-- are produced in the narrow-line region (NLR). {In the case of moderately luminous AGN, such as Seyfert galaxies, the NLR extends from $\sim$10 pc to $\sim$1 kpc\cite{Capetti96}. }

\begin{figure*}
\centering
\includegraphics[width=14cm]{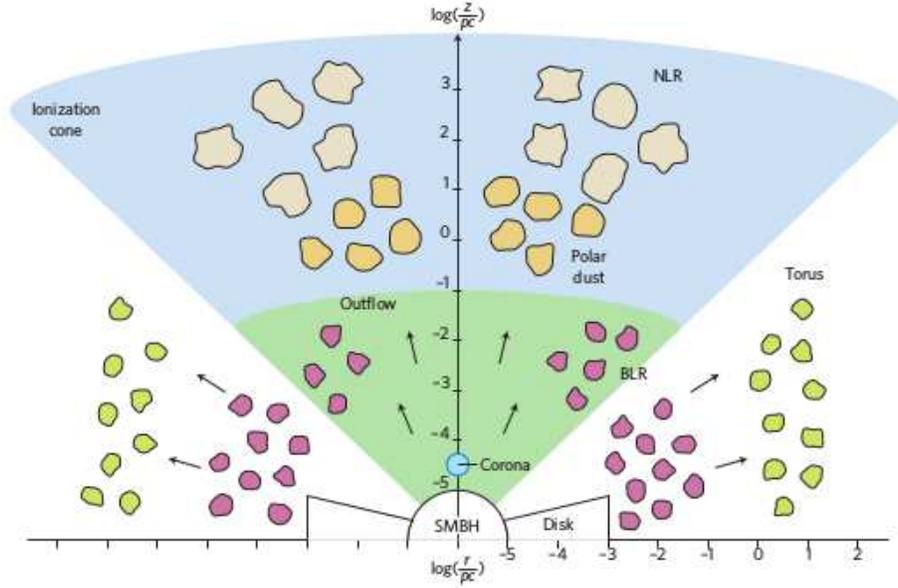}
\caption{Sketch of the main AGN structures seen along the equatorial and polar direction. From the center to host-galaxy scales:
SMBH, accretion disk and corona, BLR, torus and NLR. Different colours indicate different compositions or
densities.}
\label{fig1}
\end{figure*}

{The discovery of a highly polarized H$\alpha$ broad component in the radio galaxy 3C\,234 with position angle perpendicular to the radio axis\cite{Antonucci84} led to the development of the AGN unified model\cite{Antonucci93,Urry95}. 
These observations can be explained if the central engine is surrounded by a dusty toroidal structure, dubbed the torus, which blocks the direct emission from the BLR in type-2 AGN, and scatters the photons producing the observed polarized spectra.} This toroidal structure, of 0.1--10 pc in size {as constrained from mid-IR (MIR) imaging\cite{Packham05,Radomski08} and interferometry\cite{Burtscher13}, and more recently from sub-millimiter observations\cite{Imanishi16,Garcia16,Gallimore16}}, also collimates the AGN radiation, hence producing the bi-conical shapes of their NLRs known as ionization cones\cite{Malkan98}. 
In summary, from the very center to host-galaxy scales the main AGN structures are the accretion disk and the corona, the BLR, the torus and the NLR, as shown in Figure~\ref{fig1}.

{Another structure inferred from radio observations is the sub-pc scale maser disk: a compact concentration of clouds orbiting the SMBH and emitting in the 22 GHz maser line\cite{Greenhill96}. The maser disk is generally assumed to be co-spatial with the torus, although it is not clear whether it corresponds to its innermost part or to a geometrically thin disk which inflates in the outer part\cite{Masini16}. 
}  




{X-ray emission is ubiquitous in AGN, and is produced in a compact source located within a few gravitational radii\cite{Zoghbi:2012jk} from an accretion disk.} The study of the reprocessed and absorbed X-ray radiation can provide important information on the structure and physical properties of the circumnuclear material. 
{The level of low-ionization absorption in the X-rays} is typically parametrized in terms of the line-of-sight column density ($N_{\rm\,H}$), and AGN are considered to be obscured if $N_{\rm\,H}\geq 10^{22}\rm\,cm^{-2}$. While obscuration strongly depletes the X-ray flux at $E<10$\,keV due to the photo-electric effect, emission in the hard X-ray band ($E\gtrsim10\rm\,keV$) is less affected by obscuration. {Therefore, observations carried out using hard X-ray satellites such as {\it NuSTAR}, {\it Swift}/BAT, {\it Suzaku}/PIN, {\it INTEGRAL} IBIS/ISGRI, and {\it BeppoSAX}/PDS can probe even some of the most elusive accretion events.} Recent hard X-ray surveys have {contributed to} significantly improve our understanding of AGN obscuration, {showing that $\sim 70\%$ of all local AGN are obscured\cite{Burlon:2011dk,Ricci:2015tg}}. {While nuclear obscuration is mostly associated with dust within the torus at IR wavelengths, it can also be related to dust-free gas in the case of the X-rays.} Indeed, it is likely that X-ray obscuration is produced by multiple absorbers on various spatial scales. {This might include dust beyond the sublimation radius, and dust-free gas within the BLR and the torus\cite{Risaliti07,Maiolino:2010fu}}. This explains observations showing that, in general, the columns of material implied in the X-ray absorption are found to be comparable to or larger than those inferred from nuclear IR observations\cite{Ramos09,Burtscher16}.

Early X-ray studies revealed that, while most type-1 AGN are unobscured, type-2 AGN are usually obscured\cite{Awaki:1991rw}, supporting the unification model. {A clear example is NGC\,1068, the archetypal type-2 AGN, which has been shown to be obscured by material optically-thick to photon-electron scattering (Compton-thick or CT, $N_{\rm\,H}\geq 1.5\times 10^{24}\rm\,cm^{-2}$), which depletes most of the X-ray flux\cite{Matt:1997qy,Bauer:2015si}.} {Nevertheless for some objects with no broad optical lines no X-ray obscuration has been found\cite{Panessa:2002if}.} 
{Interestingly, many of these objects have low accretion rates, which would be unable to sustain the dynamical obscuring environment (i.e., the BLR and the torus) observed in typical AGN\cite{Nicastro:2000cq,Elitzur:2009hh}, explaining the lack of X-ray obscuration and broad optical lines.} On the other hand, studies of larger samples of objects have reported tantalizing evidence of a significant AGN population showing broad optical lines and column densities $N_{\rm\,H}\geq 10^{21.5}\rm\,cm^{-2}$ in the X-rays\cite{Merloni:2014wq}. {This has been explained considering that some obscuration is related to dust-free gas within the sublimation region associated to the BLR\cite{Davies:2015rw}.}



{The boundary between the BLR and the torus} is set by the dust sublimation temperature. The sublimation region has been resolved\cite{Kishimoto09,Weigelt12} in the near-IR (NIR) using the Very Large Telescope Interferometer (VLTI) 
and the Keck Interferometer. {From these interferometric observations it has been found that the inner torus radius scales with the AGN luminosity\cite{Kishimoto11} as $r\propto L^{1/2}$, as previously inferred from optical-to-IR time lag observations\cite{Suganuma06} and also from theoretical considerations\cite{Barvainis87}.} 

{The torus radiates the bulk of its energy at MIR wavelengths, although recent interferometry results might complicate this scenario\cite{Honig12,Honig13,Lopez16}. 
From both IR and X-ray observations it has been shown that the nuclear dust is distributed in clumps\cite{Ramos09,Markowitz:2014oq}, and further constraints on the torus 
size and geometry have been provided by MIR interferometry\cite{Burtscher13,Lopez16}. The MIR-emitting dust is compact and sometimes it appears not 
as a single component but as two or three\cite{Tristram14}.} 

\begin{figure*}
\centering
\includegraphics[width=16cm]{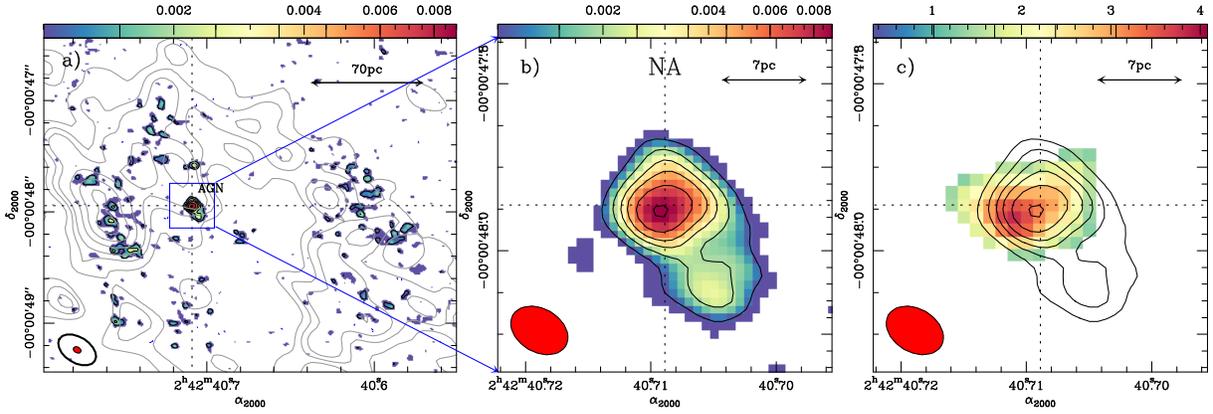}
\caption{{ALMA maps of the dust continuum and molecular gas in the nucleus of NGC\,1068\cite{Garcia16}.} (a) ALMA natural (NA)-weighted map of the dust continuum emission at 
432 $\mu$m in the circumnuclear disk of NGC\,1068. (b) Close-up of the dust continuum emission shown in the left panel. (c) Overlay of the continuum emission contours shown 
in panel (b) on the CO(6--5) emission from the torus. The red-filled ellipses at the bottom left corner of panels (b) and (c) represent the ALMA beam size at 694 GHz. The dashed
lines highlight the AGN location.}
\label{alma}
\end{figure*}

Thanks to the unprecedented angular resolution afforded by the Atacama Large Millimeter/submillimeter Array (ALMA), {recent observations have, for the first time, imaged the dust emission, the molecular gas distribution, and the kinematics from a 7--10 pc diameter disk that represents the sub-mm counterpart of the putative torus of NGC\,1068\cite{Imanishi16,Garcia16,Gallimore16} (see Figure \ref{alma}). As the sub-mm range probes the coolest dust within the torus, this molecular/dusty disk extends is twice larger than the warmer compact MIR sources detected by the VLTI in the nucleus of NGC\,1068\cite{Lopez14} and the pc-scale ionized gas and maser disks imaged in the mm regime\cite{Gallimore96,Gallimore97} which correspond to the torus innermost part. The highest angular resolution ALMA images available to date (0.07\arcsec$\times$0.05\arcsec) reveal a compact molecular gas distribution showing non-circular motions and enhanced turbulence superposed to the slow rotation pattern of the disk\cite{Garcia16}. This is confirmed by deeper ALMA observations at the same frequency\cite{Gallimore16} which permit to disentangle the low-velocity compact CO emission ($\pm$70 km~s$^{-1}$ relative to the systemic velocity) from the higher-velocity CO emission ($\pm$400 km~s$^{-1}$), which the authors interpreted as a bipolar outflow almost perpendicular to the disk.} 

Furthermore, from the left panel of Figure \ref{alma} it is clear that the torus is not an isolated structure. Instead, it is connected physically and dynamically with the circumnuclear disk (CND) of the galaxy\cite{Garcia16} ($\sim$300 pc$\times$200 pc). Indeed, previous NIR integral field spectroscopy data of NGC\,1068 revealed molecular gas streams from the CND into the nucleus\cite{Muller09}. CNDs appear to be ubiquitous in nearby AGN and they constitute the molecular gas reservoirs of accreting SMBHs\cite{Hicks13}.

\section*{Dust and gas spectral properties}


\subsection*{X-ray tracers of circunmnuclear material.}

The X-ray emitting plasma irradiates the surrounding material giving rise to several {\it reflection} features, the most important of which are the Fe\,K$\alpha$ line at 6.4\,keV and a 
Compton {\it hump} that peaks at $\sim 30\rm\,keV$\cite{Matt:1991ly}. While the Fe\,K$\alpha$ line can be produced by material with column densities as low as 
$N_{\rm\,H}\simeq 10^{21-23}\rm\,cm^{-2}$, the Compton hump can only be created by the reprocessing of X-ray photons in CT material. The Compton hump is a common feature in the X-ray spectra of AGN, showing that CT material is almost omnipresent in AGN. It is however still unclear what fraction of the Compton hump arises {from the accretion disk and what from material associated to the BLR or the torus.}

The narrow Fe\,K$\alpha$ line (FWHM$\simeq 2000\rm\,km\,s^{-1}$)\cite{Shu:2010tg} is an almost ubiquitous feature in the X-ray spectra of AGN\cite{Nandra:1994ly}, and its energy is consistent with this feature originating in lowly-ionised material\cite{Shu:2010tg}. Its origin is still under discussion, and it could be related to the torus\cite{Shu:2010tg}, to the BLR\cite{Bianchi:2008sf}, or to an intermediate region between the two\cite{Gandhi:2015zp}. {The flux of the narrow Fe\,K$\alpha$ line (compared to the intrinsic X-ray flux) is generally weaker in type-2 AGN than in type-1, and it is depleted in CT AGN with respect to less obscured AGN\cite{Ricci:2014ek}.} This would be in agreement with the idea that the circumnuclear material is axisymmetric, as predicted by the unified model, and pointing to the torus or its immediate surroundings as the region where the bulk of this line is produced. In CT material some of the Fe\,K$\alpha$ photons are bound to be down scattered, giving rise to the Compton shoulder. While the shape of this feature carries important information on the geometry and physical characteristics of the material surrounding the SMBH\cite{Matt:2002eu}, the spectral resolution of current facilities has not permitted to study it in detail.



\subsection*{Infrared tracers of circunmnuclear material.}

High angular resolution observations obtained with ground-based 8--10 m--class telescopes and with the  {\it Hubble Space Telescope} have been fundamental to characterize 
the nuclear IR spectral energy distributions (SEDs) of AGN\cite{Alonso03,Ramos09,Prieto10,Asmus14}. In general, 
while the subarcsecond resolution NIR SEDs
(1--8 \micron) of nearby type-1 AGN are {bluer} than those of type-2s, the MIR slope (8--20 \micron) is practically identical for the two types\cite{Levenson09,Prieto10,Ramos11,Asmus14}, 
indicating that the MIR emission is more isotropic than expected from a smooth torus\cite{Pier92,Pier93}. {The wavelength dependency of the IR anisotropy has been also studied 
at higher redshift using an isotropically selected sample of quasars and radio galaxies\cite{Honig11}. Longward of 12 \micron~the anisotropy is very weak, 
and the emission becomes practically isotropic at 15 \micron.} This weak MIR anisotropy explains the strong 1:1 correlation between the MIR and hard X-ray luminosities 
found for both type-1 and type-2 AGN\cite{Krabbe01,Lutz:2004gf,Asmus15}. 

Another MIR spectral characteristic used to study nuclear obscuration and commonly associated with the torus is the 9.7\,$\mu$m silicate feature. 
It generally appears in emission in type-1 AGN and in absorption in type-2 AGN, although there are exceptions\cite{Roche91,Mason09}. The amount of extinction that can be inferred from the silicate feature strength shows a correlation, although with large scatter, with the column densities derived from the X-rays\cite{Shi06}. {In general,} large columns correspond to silicate absorption and small columns to silicate emission. High angular resolution MIR spectroscopy of face-on and isolated AGN {revealed shallow silicate features in type-2 AGN\cite{Roche06,Alonso16}. A clumpy distribution of dust naturally produces these shallow absorption features, but another interpretation to explain this and other ``anomalous'' dust properties in AGN such as the reduced E$_{B-V}$/N$_H$ and A$_V$/N$_H$ ratios is a dust distribution dominated by large grains\cite{Maiolino01}.} 




\section*{Torus models}

{As a result of the small size of the torus, neither ground-based single-dish telescopes nor X-ray satellites are able to resolve it even in the most nearby AGN.}
As a consequence, different sets of IR and X-ray torus models were developed aiming to reproduce the observed SEDs and to put indirect constraints on the torus properties. Pioneering work in modelling the dusty torus\cite{Pier92,Pier93} in the IR assumed a uniform dust density distribution for the sake of simplicity. However, it was known from the very beginning that a smooth dust distribution cannot survive in the hostile AGN vicinity\cite{Krolik88}. Instead, the dust has to be arranged in dense and compact clumps. Observationally, X-ray variability studies have provided further support to a clumpy distribution of the obscuring material\cite{Markowitz:2014oq,Marinucci:2016eu}.


In order to solve the discrepancies between IR observations and the first smooth torus models (e.g. shallow silicate features, relatively {blue} IR SEDs in type-2 AGN, and small torus sizes), 
more sophisticated models were developed in the last decade. Roughly, two different sets of models can be distinguished. On the one hand, {\it physical models} aim to consider processes such as AGN and supernovae feedback, inflowing material, and disk maintenance\cite{Schartmann08,Wada02,Wada12}. On the other hand, {\it Geometrical/ad-hoc models} attempt to reproduce the IR SED by assuming a certain geometry and dust composition\cite{Nenkova08a,Nenkova08b,Honig10,Stalevski12,Siebenmorgen15}. The two types of models have advantages and disadvantages. Physical models are {potentially more realistic but it is more difficult to compare them with observations, and they generally have to assume extreme conditions to work, such as very massive star clusters or disks, or combine multiple effects like star formation, feedback and radiation with high Eddington ratios. However, much progress has been made since the first physical torus models were developed, and many of these problems are currently being solved.} On the other hand, geometrical models can be easily compared with observations but they face the problem of large degeneracies and dynamical instability. Nevertheless, much has been learned in the last years from comparing models and observations, and what is important is to be aware of the model limitations when interpreting the results\cite{Feltre12}.

\begin{figure*}
\centering
\includegraphics[width=16cm]{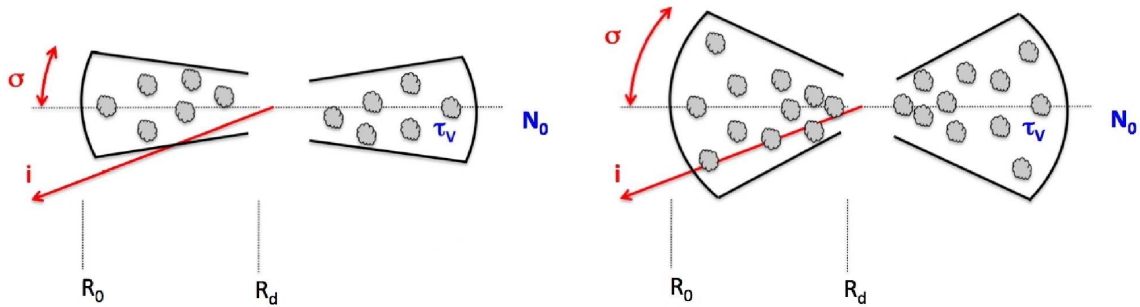}
\caption{{Sketches of two clumpy tori with different covering factors.} Smaller covering factor tori have larger photon escape probabilities associated, while larger covering factor tori are more likely to result in a type-2 AGN classification from our line of sight. Torus parameters such as the number of clumps (N$_0$), angular width ($\sigma$), optical depth of the clumps ($\tau_V$) and inclination ($i$) are labelled. R$_d$ is the dust sublimation radius and R$_o$ the outer radius of the torus.
Figure adapted from \cite{Ramos14} and based on the clumpy torus scheme\cite{Nenkova08a}.}
\label{torus}
\end{figure*}

Geometrical torus models are particularly useful for performing statistical analysis {using galaxy samples and, for example, deriving trends in the torus parameters between 
type-1 and 2 AGN. This can be done by evaluating the joint posterior distributions using Bayesian analysis\cite{Ramos11}, or a hierarchical Bayesian approach 
to derive information about the global distribution of the torus parameters for a given subgroup\cite{Ichikawa15}. Individual source fitting with geometrical 
torus models should be used when additional constraints from observables such as the ionization cones opening angle and/or orientation are considered as a priori information in the fit.} In particular, clumpy torus models have made significant progress in accounting for the IR emission of different AGN samples\cite{Mor09,Ramos09,Honig10,Alonso11,Lira13}. Examples of torus parameters that can be derived from SED modeling and compared with independent observations are e.g. the torus width ($\sigma$), which is the complement of the ionization cone half-opening angle angle; the torus outer radius (R$_o$), which can be compared with interferometry constraints; and the covering factor, which depends on the number of clumps (N$_0$) and $\sigma$ (see Figure \ref{torus}). These IR covering factors can be compared with those derived from the modeling of X-ray spectra (see next section for further details). 


It is worth noting that we do not only learn from what the models can fit, but also from what they cannot fit. For example, the NIR SEDs of some type-1 Seyferts and 
Palomar-Green quasars (PG quasars) show a $\sim$3 $\mu$m bump that cannot be reproduced with clumpy torus models only, revealing either the presence of nuclear hot dust that is not accounted for in the models or NLR flux contaminating the nuclear measurements\cite{Mor09,Alonso11}. Note, however, that in the case of the more sophisticated two-phase torus models\cite{Stalevski12}, the low-density diffuse interclump dust accounts for the NIR excess in some cases\cite{Lira13,Roseboom13}. {The NIR bump is also reproduced by recently available radiative transfer models\cite{Honig17} that assume an inflowing disk which is responsible for the NIR peak and an outflowing wind that produces the bulk of the MIR emission. Although successful in reproducing recent MIDI interferometric observations of nearby AGN\cite{Lopez16}, the number of free parameters is even larger than in clumpy torus models.}

Another example of IR SEDs that clumpy models cannot reproduce are those of {low-luminosity AGN (LLAGN) with L$_{\rm\,bol} < 10^{42}$ erg~s$^{-1}$, which show a {bluer MIR spectrum (5-35 \micron)} than those of LLAGN with L$_{\rm\,bol} \ge 10^{42}$ erg~s$^{-1}$} \cite{Gonzalez15}. This could be indicating that the torus disappears at low bolometric luminosities\cite{Elitzur:2009hh}.

In the X-rays, torus spectral models are calculated from Monte Carlo simulations of reprocessed and absorbed X-ray radiation\cite{Ikeda:2009hb,Murphy:2009hb,Brightman:2011fe,Liu:2014ff,Furui:2016qf}, and currently consider geometries more simplified than the IR models. Typical parameters obtained from these models are the column density of the torus, its covering factor and the inclination angle with respect to the system. {The two most commonly used models adopt homogeneous toroidal\cite{Murphy:2009hb,Brightman:2011fe} geometries,} and are used to study the most heavily obscured sources, for which the obscuring material acts as a sort of coronagraph, permitting to clearly observe the reprocessed X-ray radiation. These models have allowed to significantly improve the constraints on the properties of the most obscured AGN\cite{Balokovic:2014dq,Annuar:2015wd,Koss:2016fv}, and in some cases to separate the characteristics of the material responsible for the reprocessed emission and those of the obscurer\cite{Yaqoob:2012wu,Bauer:2015si}.

\section*{Covering factor of the obscuring material}
\label{covering}

The covering factor is the fraction of sky covered by the obscuring material, as seen from the accreting SMBH, and it is one of the main elements regulating the intensity of the reprocessed X-ray and IR radiation. In the last decade different trends with luminosity and redshift have been found by studying different AGN samples at different wavelengths. 

Both in the IR and X-rays the covering factor can be inferred from spectral modelling, as outlined in the previous section. 
Two additional methods are often used: i) in the IR the ratio between the MIR and the AGN bolometric luminosity is used as a proxy of the {torus reprocessing efficiency. The fraction of the optical/UV and X-ray radiation reprocessed by the torus and observed 
in the MIR is proportional to its covering factor.} ii) In the X-rays the covering factor of the gas and dust surrounding the SMBH can be estimated using a statistical argument and studying the absorption properties of large samples of AGN. The compactness of the X-ray corona implies that the value of the column density obtained from X-ray spectroscopy of single objects provides information only along an individual line-of-sight. Studying large samples of objects {allows us to probe} random inclination angles, therefore providing a better understanding of the average characteristics of the obscuring material. In fact the probability of seeing an AGN as obscured is proportional to the covering factor of the gas and dust. Therefore the fraction of sources with column density within a certain range can be used as a proxy of the mean covering factor within that $N_{\rm\,H}$ interval.

\begin{figure*}[t!]
\centering
\begin{minipage}{.49\textwidth}
\centering
\subfloat[]{\includegraphics[width=8.2cm]{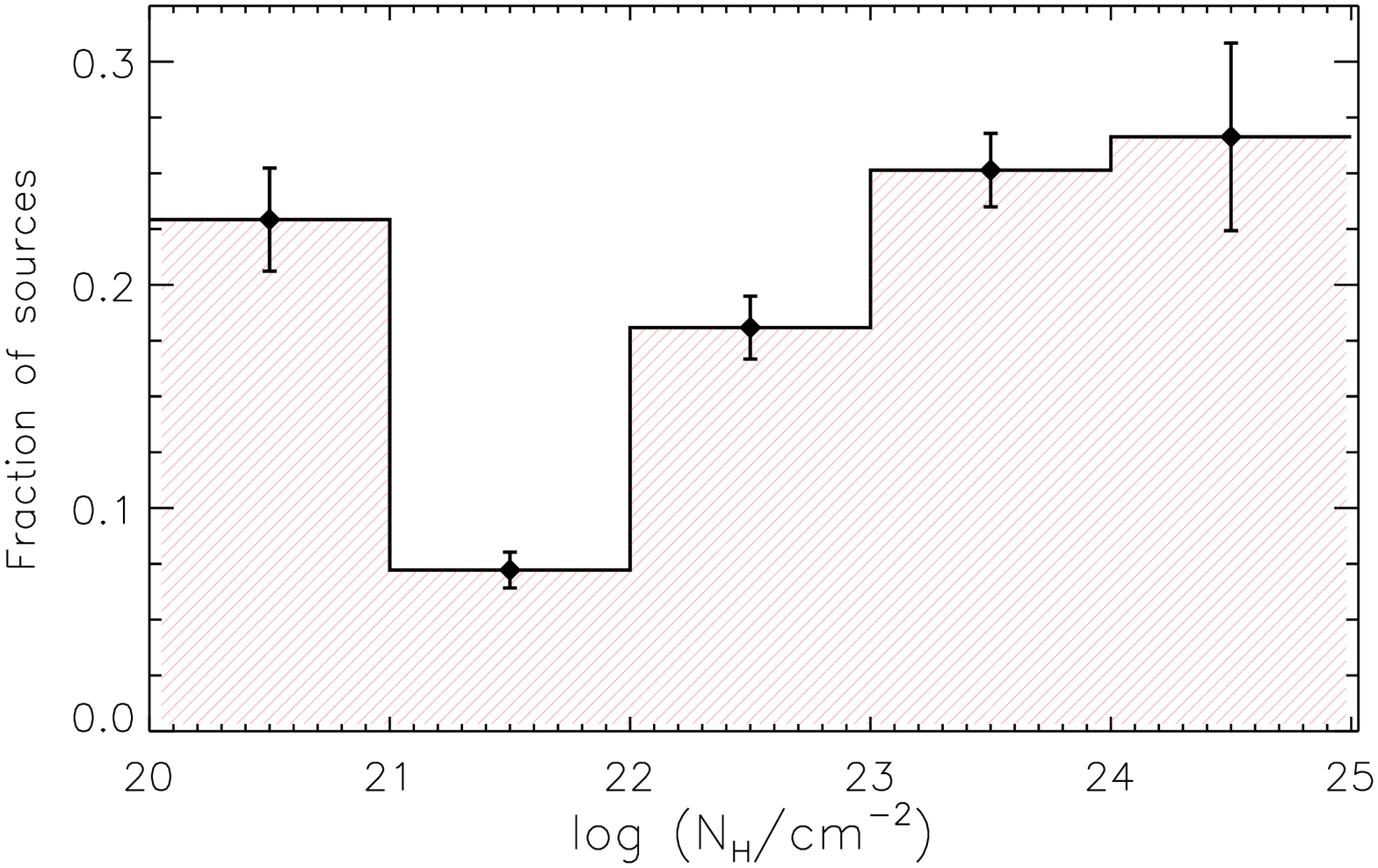}}\end{minipage}
\begin{minipage}{.49\textwidth}
\centering
\subfloat[]{\includegraphics[height=6.5cm]{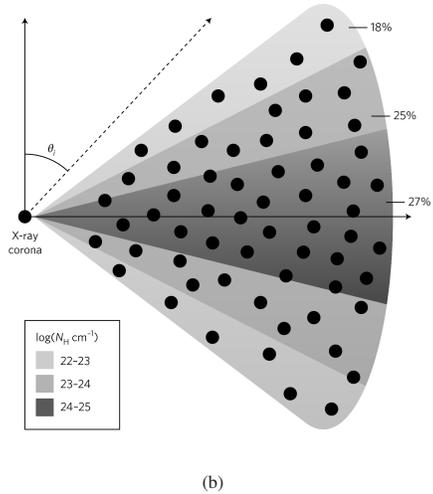}}\end{minipage}
 \begin{minipage}{1\textwidth}
  \caption{{Structure of the obscuring material in local AGN.} (a) Intrinsic column density distribution of local hard X-ray selected AGN\cite{Ricci:2015tg}, 
  showing that the average covering factor of the obscuring material is $70\%$. (b) Schematic representation of the structure of the obscuring material as inferred from 
  the intrinsic $N_{\rm\,H}$ distribution shown in the left panel. $\theta_{\rm\,i}$ is the inclination angle relative to the observer. 
    }\label{fig:NHdistribution}
     \end{minipage}
\end{figure*}

The left panel of Figure\,\ref{fig:NHdistribution} shows the intrinsic column density distribution of local AGN selected in the hard X-ray band and corrected for selection effects. Following the statistical argument outlined above, the $N_{\rm\,H}$ distribution provides important insights on the average structure of the gas and dust, and it can be used to infer the covering factors of different layers of obscuring material, assuming that the column density increases for larger inclination angles (see right panel of Figure\,\ref{fig:NHdistribution}). 
%
%
The existence of a significant population of AGN observed through CT column densities was suggested by early X-ray observations of nearby AGN\cite{Risaliti:1999dw}, which found that some of the nearest objects (Circinus, NGC\,1068, and NGC\,4945) are obscured by column densities $\geq 10^{24}\rm\,cm^{-2}$. Recent hard X-ray studies have shown that $\sim 20-27\%$ of all AGN in the local Universe are CT \cite{Burlon:2011dk,Ricci:2015tg}, and a similar percentage is found also at higher redshift \cite{Ueda:2014ix,Brightman:2014zp,Buchner:2015ve,Lanzuisi:2015qr}.

\begin{figure}[t!]
\includegraphics[width=8.5cm]{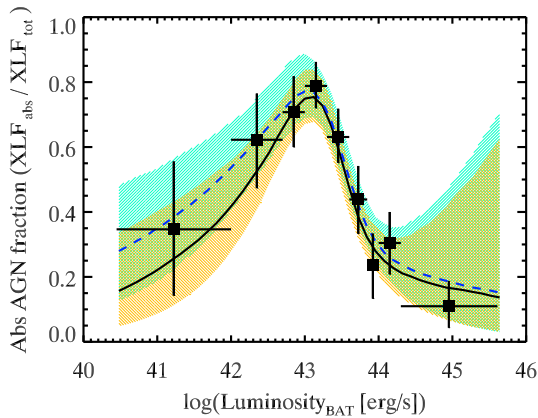}
  \caption{{Evolution of the covering factor of the obscuring material with luminosity.} Relation between the fraction of obscured Compton-thin sources ($10^{22}<N_{\rm\,H}<10^{24}\rm\,cm^{-2}$) and the 14--195\,keV luminosity for AGN detected by {\it Swift}/BAT\cite{Burlon:2011dk}.  
    }
\label{fig:fobs_L}
\end{figure}

Studies of X-ray selected samples of AGN have shown that the fraction of obscured Compton-thin ($N_{\rm\,H}=10^{22-24}\rm\,cm^{-2}$) sources decreases with luminosity\cite{La-Franca:2005kl,Akylas:2006gd,Burlon:2011dk,Merloni:2014wq,Ueda:2014ix} (see Figure\,\ref{fig:fobs_L}). A similar behavior has been observed for the Compton-thick material in a large sample of {\it Swift}/BAT AGN\cite{Ricci:2015tg}, as well as from the parameters derived from the broad-band X-ray spectroscopic analysis of a sample of local CT AGN using a physical torus model\cite{Brightman:2015fv}. 
This trend has been interpreted as being due to the decrease of the covering factor of the obscuring material with the luminosity, and it has been also reported by several studies carried out in the IR using the ratio between the MIR and the bolometric luminosity\cite{Maiolino:2007ye,Treister:2008kc,Lusso:2013vf,Stalevski:2016hl}. Furthermore, the fraction of obscured AGN at a given X-ray luminosity also increases with redshift\cite{Ueda:2014ix}, suggesting that the circumnuclear material of AGN might also evolve with Cosmic time.
Nevertheless, recent works carried out in the IR\cite{Stalevski:2016hl} and X-rays\cite{Sazonov:2015ys} have argued that if the anisotropy of the circumnuclear material is properly accounted for, the decrease of the covering factor with luminosity would be significantly weaker.


The intensity of the reprocessed X-ray radiation depends on the covering factor of the obscuring material, which would lead to expect a direct connection between reflection features and X-ray luminosity. While the relation between the Compton hump and the luminosity is still unclear, a decrease of the equivalent width of the Fe\,K$\alpha$ line with the luminosity (i.e. the X-ray Baldwin effect\cite{Iwasawa:1993ez}) has been observed in both unobscured\cite{Bianchi:2007os,Shu:2010tg} and obscured\cite{Ricci:2014ek} AGN. Moreover, the slope of the X-ray Baldwin effect can be reproduced by the relation between the covering factor and the luminosity found by X-ray studies\cite{Ricci:2013hi}, also suggesting a relation between the two trends.

%


The relation between the obscuring material and the AGN luminosity has been often explained as being a form of {\it feedback}, with the radiation field of the AGN 
cleaning out its circumnuclear environment\cite{Fabian:2006sp}. 
Interestingly, the decrease of the covering factor with luminosity does not extend to the highest bolometric luminosities ($10^{46-48}\rm\,erg\,s^{-1}$), 
where about half of the AGN population seem to be obscured\cite{Assef:2015ly}. We note that these obscured AGN are not necessarily type-2 AGN in the optical range.
In the low-luminosity regime, evidence for a decrease of the fraction of obscured sources for X-ray luminosities $<10^{42}\rm\,erg\,s^{-1}$ has been 
observed\cite{Burlon:2011dk} (Figure\,\ref{fig:fobs_L}). This result is supported by MIR observations of nearby AGN, which claim that the torus disappears 
at luminosities below the before-mentioned limit\cite{Gonzalez15}. As discussed in previous sections, these results can be explained if low-luminosity AGN 
fail to sustain the AGN internal structures\cite{Elitzur:2006ec}. {This idea would also explain the observed decrease of the Fe\,K$\alpha$ intensity with 
respect to the X-ray flux at low-luminosities found using {\it Suzaku}\cite{Kawamuro:2016lq}.} 

The modelling of IR SEDs has also provided important insights on the unification scheme. {Using clumpy torus models, it has been claimed} that the covering factors of type-2 
tori are larger (i.e., more clumps and broader tori) than those of type-1 AGN\cite{Ramos11,Ichikawa15,Alonso11,Mateos16}. 
{This would imply,} first, that the observed differences between type-1 and type-2 AGN are not due to orientation effects only, as proposed in the simplest unification model, but also to {the dust covering factor}. 
Second, that the torus is not identical for all AGN of the same luminosity. Therefore, the classification of an AGN as a type-1 or type-2 is probabilistic\cite{Elitzur12} (see Figure \ref{torus}). 
The larger the covering factor, the smaller the escape probability of an AGN-produced photon. Although these results are model-dependent, they reflect the observed differences between the IR 
SEDs of type-1 and type-2 AGN. It is noteworthy that in the radio-loud regime it has been found that the radio core dominance parameter (the ratio of pc-scale jets to radio lobes emission) 
agrees with the optical classification of a subsample of 3CR radio galaxies at z$\geq$1\cite{Marin2016}. This would be indicating a small dispersion of torus opening angle and inclination for 
luminous type-1 and type-2 radio-loud AGN. Unfortunately, this analysis is not possible in radio-quiet AGN.

While models reproducing at the same time both the X-ray and MIR spectral properties of the reprocessed radiation are still missing, the values of the covering factors obtained by using 
MIR\cite{Alonso11,Lira13,Ichikawa15} and X-ray torus models are consistent\cite{Brightman:2015hb}, albeit with large uncertainties and for a handful of AGN only.

\section*{Variability of the line-of-sight obscuring material}
\label{variability}

\begin{figure*}[t!]
\centering
\subfloat[]{\includegraphics[width=6.9cm]{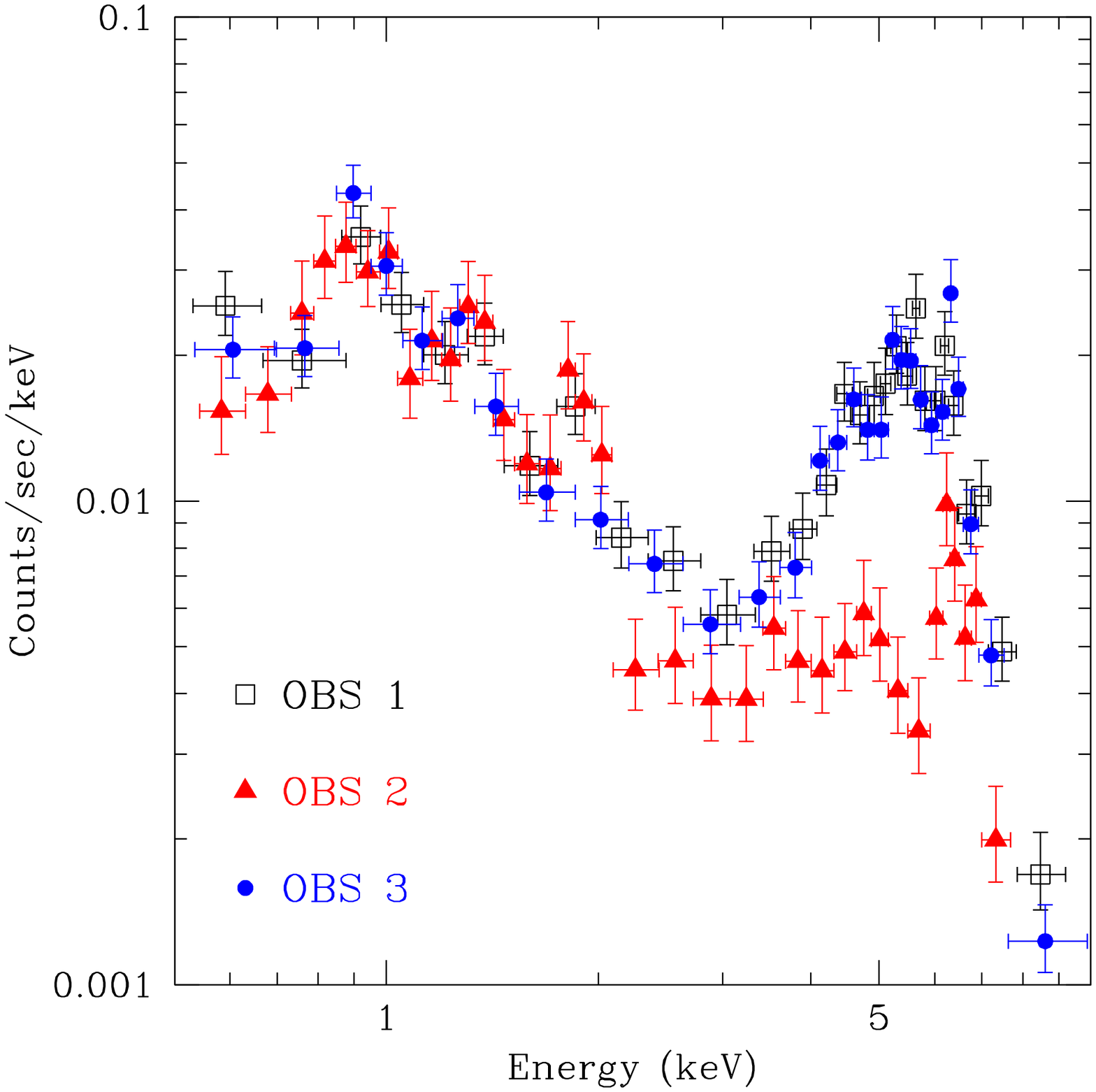}}
\subfloat[]{\includegraphics[width=8.6cm]{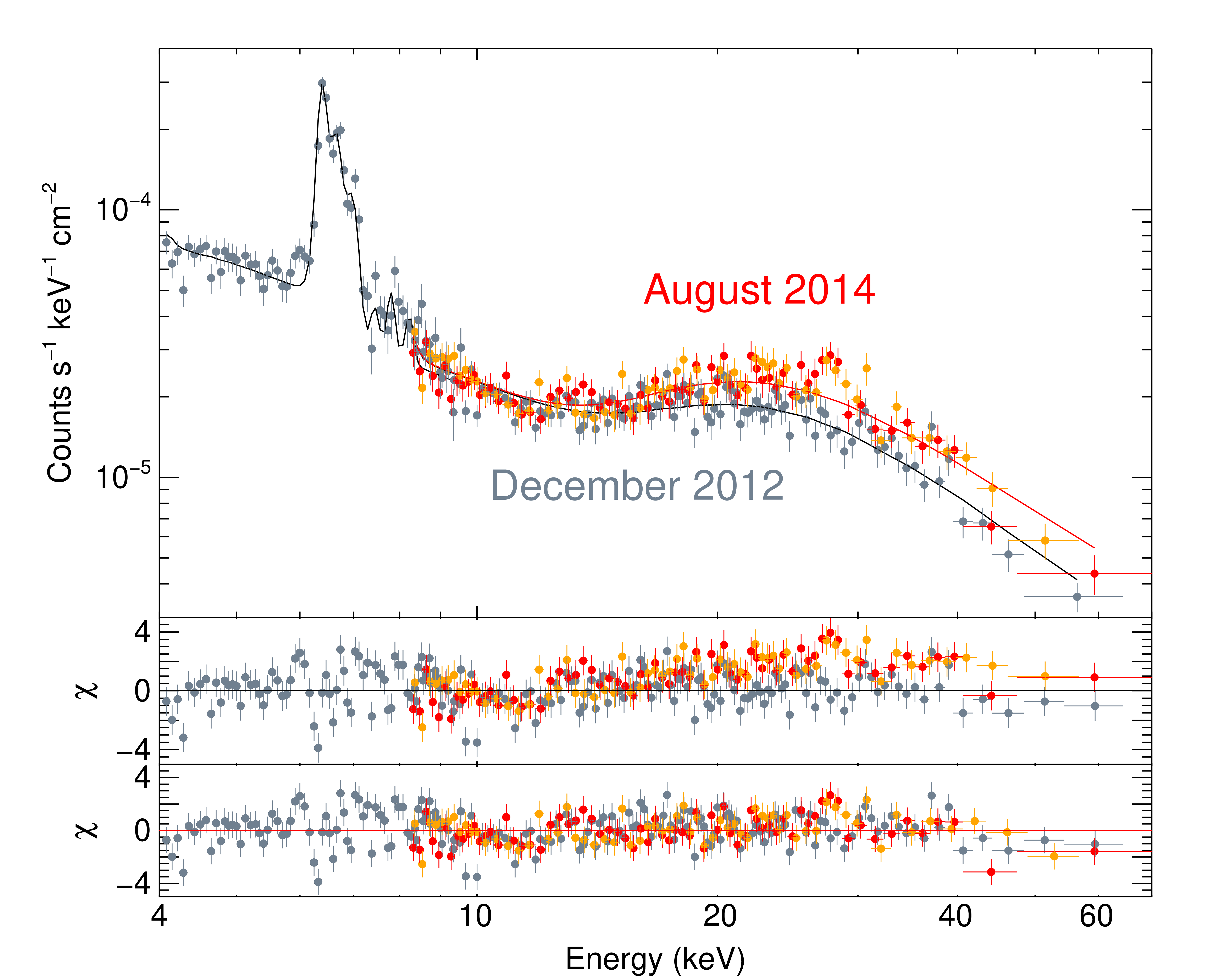}}
\caption{{Examples of absorption variability in the X-ray spectrum of nearby AGN.} (a) Absorption variability observed in the X-ray spectrum of NGC\,1365\cite{Risaliti07}. The three {\it Chandra} observations were carried out two days from each other. The spectrum of the second observation clearly shows the flat continuum and prominent Fe\,K$\alpha$ line typical of heavily obscured AGN. (b) {\it NuSTAR} observations of NGC\,1068\cite{Marinucci:2016eu}, showing an excess above 20\,keV in August 2014 that can be explained by a cloud moving away from the line-of-sight, allowing for the first time to observe the nuclear X-ray continuum of this source.}
\label{fig:variableNH}
\end{figure*}

Studies carried out in the X-rays have found variations of the column densities of the obscuring material for several dozens AGN, confirming the idea that the obscuring material is clumpy and not homogeneous, and very dynamic. In about ten objects\cite{Risaliti07,Guainazzi:2002mz,Piconcelli:2007bh}, these variations were found to be very extreme, with the line-of-sight obscuration going from CT to Compton-thin (and vice versa) on timescales of hours to weeks (see left panel of Figure\,\ref{fig:variableNH}). This is consistent with the absorber originating in the BLR. Due to the strong changes in their spectral shapes, these objects are called {\it changing-look AGN}. For the archetypal of these objects, NGC\,1365, it has been found that the obscuring clouds have a cometary shape\cite{Maiolino:2010fu}, with a high-density head and an elongated structure with a lower density. Even objects such as Mrk\,766\cite{Risaliti:2011jl}, which are usually unobscured, have been found to show eclipses produced by highly-obscuring material. For this object, highly-ionized blueshifted iron absorption lines (Fe\,XXV and Fe\,XXVI) were also detected, showing that the absorbing medium is outflowing with velocities ranging from 3,000 to 15,000\,$\rm\,km\,s^{-1}$. Interestingly, due to the mass loss from the cometary tail, clouds would be expected to be destroyed within a few months\cite{Maiolino:2010fu}. This suggests that the BLR must be very dynamic, with gas clouds being created and dissipating continuously. The origin of these clouds is still unclear, but it has been suggested that they might be created in the accretion disk\cite{Elitzur:2009hh}.

Evidence for a clumpy obscurer on scales larger than the BLR have also been found. 
A study carried out using the {\it Rossi X-ray Timing Explorer} has recently discovered variation of the absorbing material (with $N_{\rm\,H}\sim 10^{22-23}\rm\,cm^{-2}$) on timescales of months to years  for several objects\cite{Markowitz:2014oq}. For seven AGN the distance of the obscuring clouds locates them between the outer side of the BLR and up to ten times the distance of the BLR, suggesting that they are associated with clumps in the torus. Recent {\it NuSTAR} observations of NGC\,1068\cite{Marinucci:2016eu} found a $\sim$30\% flux excess above 20\,keV in August 2014 with respect to previous observations (right panel of Figure\,\ref{fig:variableNH}). The lack of variability at lower energies permitted to conclude that the transient excess was due to a temporary decrease of the column density, caused by a clump moving out of the line-of-sight, which enabled to observe the primary X-ray emission for the first time. {In the MIR, results from dust reverberation campaigns using data from the {\it Spitzer Space Telescope} are consistent with the presence of clumps located in the inner wall of the torus\cite{Vazquez15}}.


Therefore, we know from absorption variability studies that both the BLR and the torus are not homogeneous structures, but clumpy and dynamic regions which might be generated as part of an outflowing wind. 

\section*{Gas and dust in the polar region}

\begin{figure*}
\centering
\subfloat[]{\includegraphics[width=7.7cm]{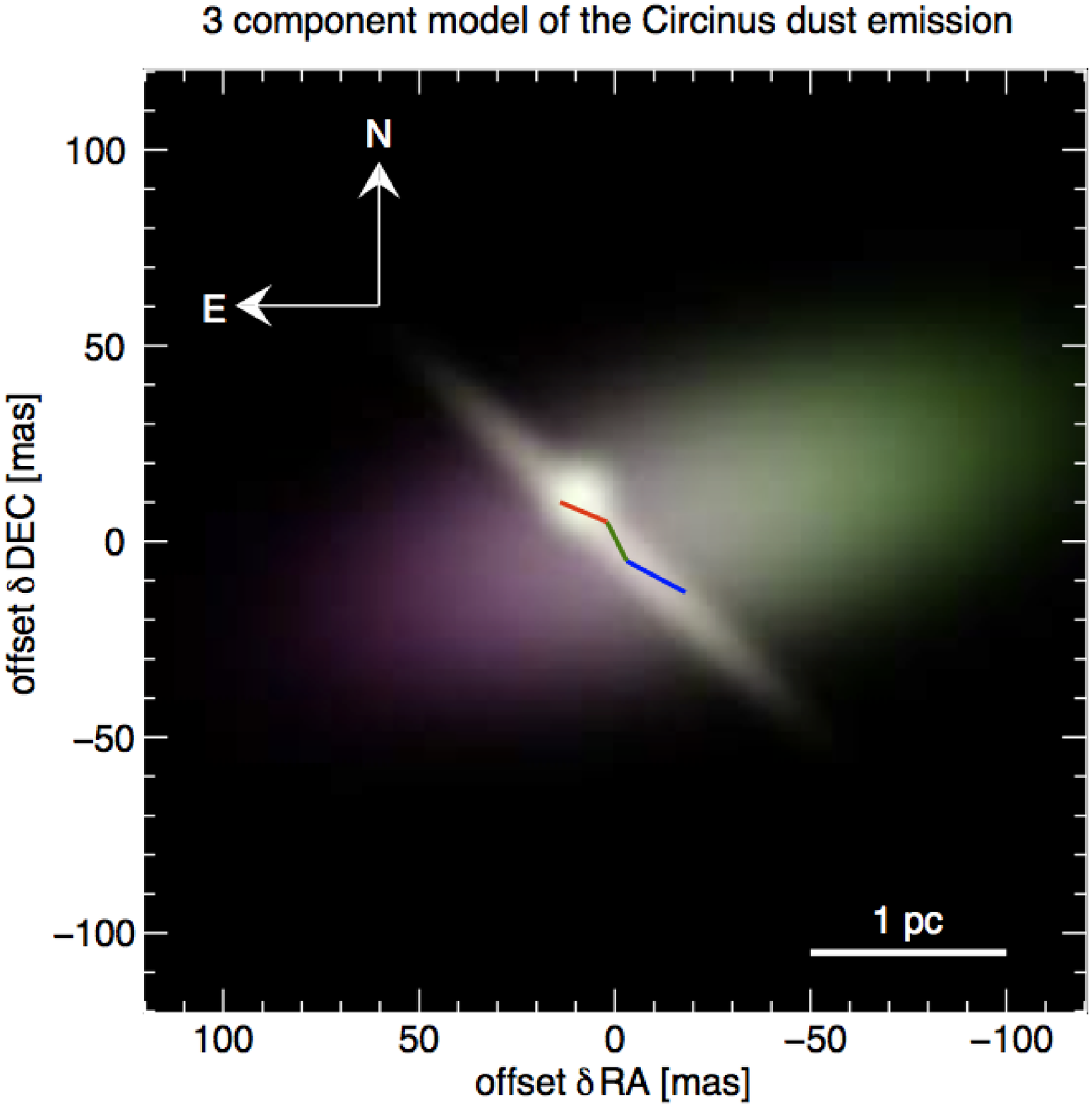}}
\subfloat[]{\includegraphics[width=7.9cm]{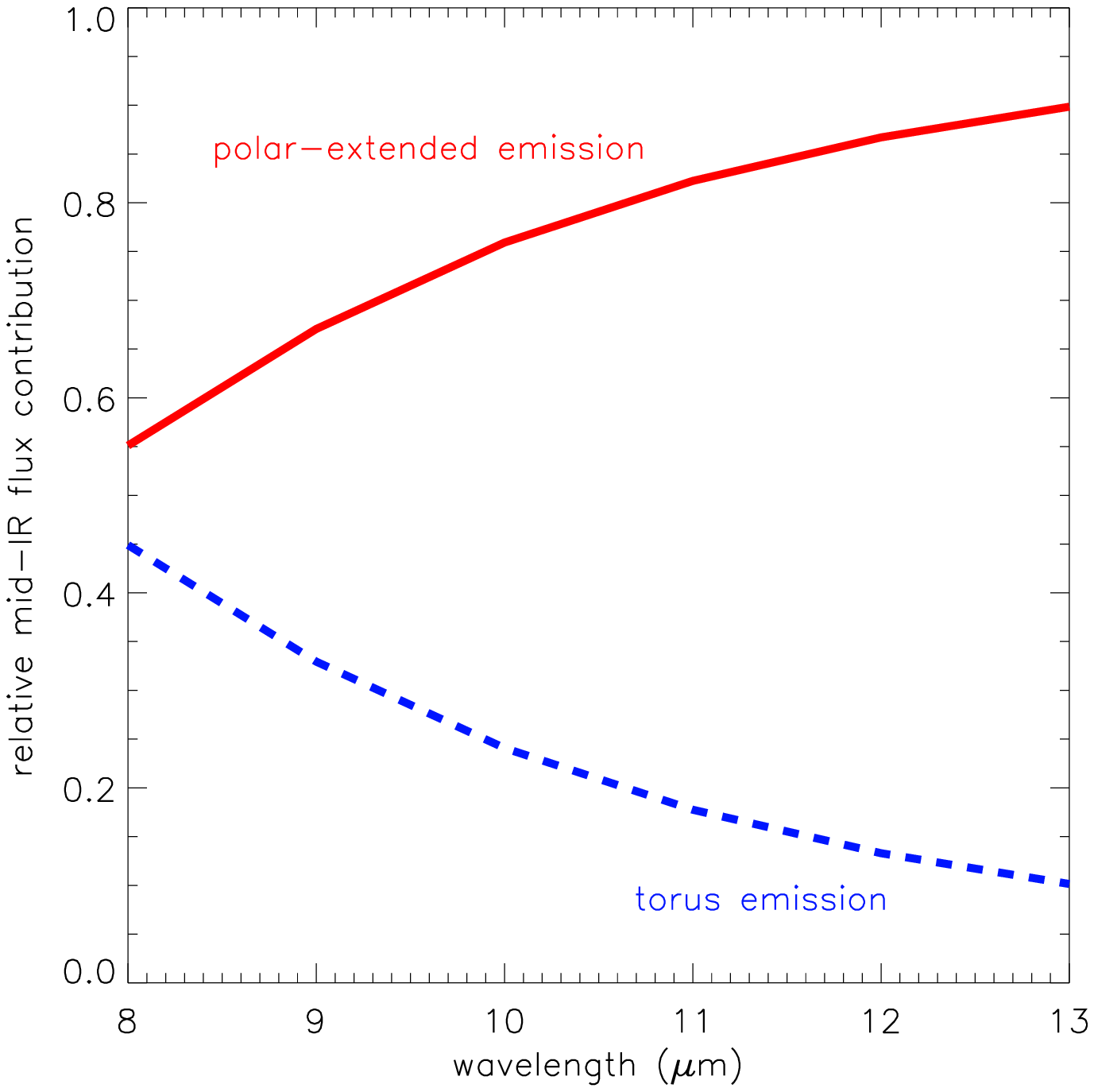}}
\caption{{Interferometry results from VLTI/MIDI observations of two nearby AGN.} (a) 3-component model that reproduces the observed MIDI visibilities obtained for the Circinus galaxy\cite{Tristram14}. 
This includes a polar component containing most of the MIR emission, a disk component, and 
an unresolved component. (b) Relative contributions of the disk and polar components derived from the model that better reproduces the MIDI observations of the type-1 AGN NGC\,3783\cite{Honig13}.}
\label{polar}
\end{figure*}

In the last decade, MIR interferometry has represented a major step forward in the characterization of nuclear dust in nearby AGN. VLTI/MIDI interferometry of 23 AGN has revealed that a large part of the MIR flux is concentrated on scales between 0.1 and 10 pc\cite{Tristram09,Burtscher13}. Besides, for the majority of the galaxies studied, two model components are needed to explain the observations\cite{Burtscher13}, instead of a single toroidal/disk structure. Moreover, for some of these sources one of these two components appears elongated in the polar direction. Detailed studies of four individual sources performed with MIDI\cite{Honig12,Honig13,Tristram14,Lopez14} have shown further evidence for this nuclear polar component (see Figure \ref{polar}). More recently, a search for polar dust in the MIDI sample of 23 AGN has been carried out\cite{Lopez16}, and this feature has been found in one more galaxy. Thus, up to date, evidence for a diffuse MIR-emitting polar component has been found in five AGN, including both type-1 and type-2 sources (Circinus, NGC\,424, NGC\,1068, NGC\,3783, and NGC\,5506). This polar component appears to be brighter in the MIR than the more compact equatorial structure, and {it has been interpreted as an outflowing dusty wind driven by radiation pressure\cite{Honig12}. Indeed, radiation-driven hydrodynamical models\cite{Wada12,Wada16} taking into account both AGN and supernovae feedback can reproduce geometrically thick pc-scale disks and also polar emission, although these features are rather transient in nature even when averaged over time. 
} 

It is noteworthy that this polar component was first detected in NGC\,1068 using high angular resolution MIR observations ($\le$0.5\arcsec) obtained with {single-dish 4--10 m-class telescopes\cite{Cameron93,Bock00,Mason06}. These observations revealed that
the point source was only responsible for $\sim$30--40\% of the 8--24.5 \micron~emission, and the remaining 60--70\% was emitted by dust in the ionization cones. Thus, the bulk of the nuclear MIR flux comes from polar dust within the central 70 pc of NGC\,1068\cite{Mason06}.} More recently, a similar result has been reported for 18 active galaxies observed with 8 m-class telescopes in the MIR\cite{Asmus16}. The resolved emission is
elongated in the polar direction (i.e. NLR dust), it represents at least 40\% of the MIR flux, and it scales with the [O IV] flux. This is in line with the results from MIR interferometry\cite{Honig12,Honig13}, which indicate that the bulk of the MIR emission comes from a diffuse polar component, while the NIR flux would be dominated by a compact disk (see right panel of Figure \ref{polar}). The difference between single-dish telescopes and MIR interferometry studies is the scale of the IR-emitting regions probed.

{The existence of non-nuclear reflecting material in obscured AGN has been confirmed in the X-ray regime by {\it Chandra} studies of the Fe\,K$\alpha$ line, which showed that part of the emission originates from an extended region.} This has been found for some of the nearest heavily obscured AGN, such as NGC\,1068\cite{Young:2001jw}, for which $\sim$30\% of the Fe\,K$\alpha$ emission has been found to originate in material located at $\gtrsim$140 pc\cite{Bauer:2015si} and it seems to be aligned with the NLR. {Similarly, the 0.3--2\,keV radiation has also been found to coincide with the NLR in obscured AGN\cite{Bianchi:2006kq}.}

If proved to be common features in a significant fraction of AGN, co-existing compact disks/tori and polar dust components {should be incorporated in the models\cite{Honig17}} and could explain the observed NIR and MIR bumps seen in the SEDs of some type-1\cite{Mor09,Alonso11,Ichikawa15} and type-2 AGN\cite{Lira13}. Besides, the polar emission has been proposed as an alternative scenario to explain {the weak MIR anisotropy observed in active galaxies\cite{Honig11},} which is responsible for the strong 1:1 X-ray/MIR correlation slopes found for type-1 and type-2 AGN\cite{Gandhi:2009pd,Ichikawa:2012uo,Asmus15}. However, as explained in previous sections, a toroidal clumpy distribution also explains the weak MIR anisotropy\cite{Levenson09} and more sophisticated clumpy models account for the NIR excess of the nuclear SEDs, {either including a polar component in addition to the torus\cite{Honig17} or not\cite{Stalevski12}.}

\section*{Current picture and the future of IR and X-ray studies of nuclear obscuration in AGN}

In the past 10--15 years, studies of AGN in the IR and X-rays have provided important information on the characteristics of the nuclear environment of accreting SMBHs, 
showing that its nature is extremely complex and dynamic. {The obscuring structure is is compact, clumpy, and not isolated, but connected with the host galaxy via gas inflow/outflows.} 
From the IR point of view, it is a transition zone between the dust-free BLR clouds and the NLR and, at least in some galaxies, it consists of two structures: an equatorial disk/torus 
and a polar component. This polar component would be part of the outflowing dusty wind predicted by radiation-driven hydrodynamical models. In the case of the X-rays, the obscuration 
is produced by multiple absorbers on various spatial scales, but mostly associated with the torus and the BLR. The covering factor of the obscuring material depends on the luminosity 
of the system and possibly on the redshift, and it is important to take these dependencies into account to explain observations of both high- and low-luminosity AGN. The covering 
factor should also be considered in our current view of AGN unification, as the classification of an AGN as type-1 or type-2 does not depend on orientation only, but also on the 
AGN-produced-photon escape probability.

In the next decade the new generation of IR and X-ray facilities will {contribute greatly to our} understanding of the structure and physical properties of the nuclear material, and to shed light on relevant open question such as: what is the relationship between the physical parameters of the accreting system and the circumnuclear material? Are the torus and BLR produced by outflows in the accretion disk? Is the polar dust ubiquitous?, and how much does it contribute to the IR emission?

{In the X-ray regime, {\it NuSTAR}, {\it XMM-Newton}, {\it Chandra}, {\it Swift} and {\it INTEGRAL} will continue carrying out broad-band X-ray observations of AGN, providing tighter constraints on the most obscured accretion events and on the characteristics of the circumnuclear material through studies of the reprocessed X-ray radiation.} The recently launched X-ray satellite {\it ASTROSAT}\cite{Singh:2014pd}, thanks to its large effective area and broad band X-ray coverage, will be ideal to study absorption variability, and will improve our understanding of the properties of the BLR clouds. {\it eROSITA} (\href{http://adsabs.harvard.edu/abs/2012arXiv1209.3114M}{Merloni et al. 2012}), on board the {\it Spectrum-Roentgen-Gamma} satellite, will carry out a deep survey of the entire X-ray sky in the 0.5--10\,keV range, and is expected to detect tens of thousands of obscured AGN. {This will certainly improve} our understanding of the relation between obscuration and the accretion and host galaxy properties.

On longer timescales, {\it Athena} (\href{http://adsabs.harvard.edu/abs/2013arXiv1306.2307N}{Nandra et al. 2013}), and before that the successor of {\it Hitomi}, will enable studies of reflection features in AGN with an exquisite level of detail, exploiting the energy resolution of a few eV of micro-calorimeters. High-resolution spectroscopy studies of AGN {will make possible to disentangle} the different components (arising in the BLR, torus, NLR) of the Fe\,K$\alpha$ line, and to set tighter constraints on the properties of the circumnuclear material using the Compton shoulder\cite{Odaka:2016fv}.
NASA recently selected the {\it Imaging X-ray Polarimetry Explorer}\cite{Weisskopf:2016qd} (IXPE) mission to be launched in the next decade. X-ray polarimetry will open a new window in the study of the close environment of AGN, since the reprocessed X-ray radiation is bound to be polarised. 

To date, IR interferometry has provided constraints on the size and distribution of nuclear dust for about 40 AGN. Now, a second generation of interferometers for the VLTI are coming online. In the NIR, GRAVITY\cite{Eisenhauer11} will be able to observe $\sim$20 nearby AGN with unprecedented sensitivity and high spectral resolution, allowing to estimate reliable SMBH masses and put constraints on the geometry of the BLR. In the MIR, MATISSE\cite{Lopez14b} will combine the beams of up to 4 VLTI telescopes {to produce images that will serve to analyze} the dust emission at 300--1500 K in the central 0.1-–5 pc of the closest AGN. In the NIR and MIR, the {\it James Webb Space Telescope\cite{Gardner2006}} ({\it JWST}) will represent a revolution in terms of sensitivity and wavelength coverage. 
Faint low-luminosity and high-redshift AGN will be accessible at subarcsecond resolution from 0.6 to 28 \micron~for the first time. 
Finally, in the sub-mm regime ALMA will continue providing the first images of the nuclear obscurer in nearby AGN. In Cycle 4 and later, {ALMA will fully resolve} the gas kinematics 
from galaxy scales to the area of influence of the SMBH in nearby AGN. This will serve to characterize the inflowing/outflowing material in the nucleus and its connection with the 
host galaxy, leading to a better understanding of the feeding/feedback mechanisms in AGN.

With the advent of all the facilities described above, in order to fully exploit the wealth of data that will be available in the next decade, it will be necessary for the community 
to develop physical AGN spectral models that could self-consistently reproduce reprocessed X-ray radiation and MIR emission, ideally considering polarization as well. 

\section*{Acknowledgements}

The authors acknowledge Almudena Alonso-Herrero, Poshak Gandhi, Nancy A. Levenson, Marko Stalevski and the referees for useful comments that helped to improve this Review. 
CRA acknowledges the Ram\'on y Cajal Program of the Spanish Ministry of Economy and Competitiveness through project RYC-2014-15779 and the Spanish Plan Nacional de Astronom\' ia y 
Astrof\' isica under grant AYA2016-76682-C3-2-P. CR acknowledges financial support from the China-CONICYT fellowship program, FONDECYT 1141218 and Basal-CATA PFB--06/2007.
This work is sponsored by the Chinese 
Academy of Sciences (CAS), through a grant to the CAS South America Center for Astronomy (CASSACA) in Santiago, Chile. \\

\noindent{Correspondence should be addressed to the two authors.}

\section*{Author contributions}
The two authors contributed equally to this work. They both decided the concept of the Review and provided/adapted the figures that appear on it. 
CR and CRA wrote the X-ray and IR part of the text, respectively, and worked together to put them in common.  


%
%
%

\end{document}